\documentclass[conference]{IEEEtran}
\usepackage{cite}
\usepackage{amsmath,amssymb,amsfonts,amsthm}
\usepackage{algorithmic}
\usepackage{algorithm}
\usepackage{enumerate}  
\DeclareMathOperator{\Tr}{Tr}
\usepackage{comment}
\usepackage[english]{babel}

\newtheorem{lemma}{\quad \textit{Lemma}}

\usepackage{subcaption}
\usepackage{graphicx}
\usepackage{textcomp}
\usepackage{optidef}
\usepackage{xcolor}

\begin{document}

\title{Joint Semantic Communication and Target Sensing for 6G Communication System}
\author{
\IEEEauthorblockN{Yinchao Yang\IEEEauthorrefmark{1},
Mohammad Shikh-Bahaei\IEEEauthorrefmark{1},
 Zhaohui Yang\IEEEauthorrefmark{2}\IEEEauthorrefmark{3},
Chongwen Huang\IEEEauthorrefmark{2}\IEEEauthorrefmark{3},
         Wei Xu\IEEEauthorrefmark{4},
                  \\ and Zhaoyang Zhang\IEEEauthorrefmark{2}\IEEEauthorrefmark{3}
                 }
	\IEEEauthorblockA{
 		$\IEEEauthorrefmark{1}$Department of Engineering 
 King's College London 
London, UK \\
			$\IEEEauthorrefmark{2}$College of Information Science and Electronic Engineering, Zhejiang University, Hangzhou, China\\
   	$\IEEEauthorrefmark{3}$Zhejiang Provincial Key Laboratory of Info. Proc., Commun. \& Netw. (IPCAN), Hangzhou, China\\
                $\IEEEauthorrefmark{4}$National Mobile Communications Research Laboratory, Southeast University, Nanjing, China\\
			E-mails: 
 yinchao.yang@kcl.ac.uk,
   m.sbahaei@kcl.ac.uk,
   yang\_zhaohui@zju.edu.cn,
 chongwenhuang@zju.edu.cn,\\
 Wxu@seu.edu.cn, 
   ning\_ming@zju.edu.cn
		}}

%\author{\IEEEauthorblockN{Yinchao Yang}
%\IEEEauthorblockA{\textit{Department of Engineering} \\
%\textit{King's College London}\\
%London, UK \\
%yinchao.yang@kcl.ac.uk}
%\and
%\IEEEauthorblockN{Mohammad Shikh-Bahaei}
%\IEEEauthorblockA{\textit{Department of %Engineering} \\
%\textit{King's College London}\\
%London, UK \\
%m.sbahaei@kcl.ac.uk}
%}
 
%\and
%\IEEEauthorblockN{Yao Sun}
%\IEEEauthorblockA{\textit{James Watt School of Engineering} \\
%\textit{The University of Glasgow}\\
%Glasgow, UK \\
%Yao.Sun@glasgow.ac.uk}

\maketitle

 \begin{abstract}
  This paper investigates the secure resource allocation for a downlink integrated sensing and communication system with multiple legal users and potential eavesdroppers. In the considered model, the base station (BS) simultaneously transmits sensing and communication signals through beamforming design, where the sensing signals can be viewed as artificial noise to enhance the security of communication signals. To further enhance the security in the semantic layer, the semantic information is extracted from the original information before transmission. The user side can only successfully recover the received information with the help of the knowledge base shared with the BS, which is stored in advance. Our aim is to maximize the sum semantic secrecy rate of all users while maintaining the minimum quality of service for each user and guaranteeing overall sensing performance. To solve this sum semantic secrecy rate maximization problem, an iterative algorithm is proposed using the alternating optimization method. The simulation results demonstrate the superiority of the proposed algorithm in terms of secure semantic communication and reliable detection. 

   %  This paper explores the integration of sensing and semantic extraction for secure communication within a multi-user and multi-eavesdropper environment. To provide physical layer security for the confidential message, sensing signals are manipulated as artificial noise. On the semantic layer, the confidential message can be protected through semantic extraction. In the considered model, we aim to maximize the overall worst-case semantic secrecy rate of all communication users while maintaining the minimum quality of service for each user and guaranteeing overall sensing performance. We propose an iterative algorithm to solve this problem using the alternating optimization method. The simulation results demonstrate the superiority of the proposed algorithm in terms of secure semantic communication and reliable detection. 
 \end{abstract}

\section{Introduction}
The bottleneck of spectrum shortage is brought on by the proliferation of emerging applications with high data-rate requirements, such as virtual reality, holographic communication, and high-mobility sensing \cite{8468002,zhang2019neural, nehra2010spectral, bobarshad2009m, zhang2020ensemble, ding2023partial, 10293761, 9120678, 9417573,10097158, 10095827}. To meet the strict requirements in emerging applications, three functions must be synchronized, i.e., mobile sensing, communication, and computation. While these three functions operate independently in conventional wireless networks, they cannot meet the demands of the sixth-generation (6G) communication system including immersive communication, hyper-reliable and low latency communications, integrated sensing and communication, and integrated AI and communication. 
%, which are: ultra-reliable and low latency communications, Enhanced Mobile Broadband and Massive Machine-Type Communications.

It is therefore imperative that academics and industry work together to develop novel approaches to this problem. A rising technology, integrated sensing and communication (ISAC), has recently gained attention. Researchers initially focused on sharing the spectrum between communication and sensing. It was called communication and radar spectrum sharing \cite{liu2017robust}. Thereafter, further designs have been proposed where the spectrum and hardware infrastructure are both shared between communication and sensing \cite{liu2018toward}. As a result of its dual functions, ISAC significantly increases spectrum efficiency as well as supports emerging applications, which will play an important role in future networks.

With the rapid development of artificial intelligence, semantic communication has also proven to be an effective technology for addressing resource scarcity \cite{qin2021semantic, xu2023edge, gunduz2022beyond, chen2023big}. A breakthrough in semantic communication is that it transcends Shannon's paradigm, as opposed to conventional communication, which is constrained by Shannon's capacity limit. Rather than transmitting the entire message, semantic communication focuses on extracting and transmitting just its meaning. Several semantic coding strategies have been proposed for different data formats. For example, the work in \cite{xie2021deep} proposed machine learning approaches for processing text messages. Moreover, the authors in \cite{10123081, lyu2023semantic} concentrated on machine learning applications for image semantic coding.

To achieve efficient transmission, semantic communication works in the following steps. In the first step, the transmitter extracts the meaning of the source messages (e.g., text, image, video) using its own knowledge base. The transmitter's knowledge base is built on common knowledge (shared between both the transmitter and the receiver) and private knowledge. With the help of autoencoders \cite{yang2023secure}, this step can be achieved. In the second step, semantic messages are then transmitted over wireless channels. In the third step, the receiver recovers the meaning of the original messages by utilizing its local knowledge base. %which includes both common and private knowledge. 
However, the above semantic communication works \cite{xie2021deep,10123081,lyu2023semantic,yang2023secure} did not consider the integration of sensing and semantic communication, even though the joint design of sensing and semantic resource can further enhance the performance of the system.

In this paper, we investigate the 
integrated sensing and semantic communication (ISSC) model, which embeds semantic communication into the ISAC model, not only making more spectrum resources available but also adding an extra layer of security. The ISSC model differs from traditional physical layer security \cite{10013039}, in that, even if an eavesdropper intercepts a message, he or she may not be able to understand its meaning in the semantic layer. 
To the best of our knowledge, there is no existing work that focuses on joint semantic communication and sensing. 
The main contributions of this paper are as follows:
\begin{enumerate}
    \item  We propose the framework of an integrated sensing and semantic communication system as a spectrum-efficient and secure method. To ensure the security of the system, the sensing signal can be viewed as artificial noise for the communication signal and the semantic communication technique is used to protect the meaning of the transmitted communication signal. 

    \item In order to determine the explicit performance indicator of semantic communication, we take into account the well-known BLEU score.
    The semantic rate expression is obtained through considering the effect of both semantic compression ratio and semantic computation power.
    Based on the considered model, the joint sensing, communication, and semantic computation optimization problem is formulated by maximizing the sum secure semantic rate. To solve this problem, an alternating algorithm is proposed with iteratively solving the sensing beamforming, communication beamforming, and semantic compression ratio. 
    %In addition, we propose the idea of using a natural logarithm function to compute the computation power for extracting a semantic message. 
\end{enumerate} 

\begin{comment}
\subsection*{List of Notations:}
Capital boldface letters are used to represent matrices and lowercase boldface letters are used to represent vectors. Scalars are represented by lower and upper-case normal fonts. $\mathbb{C}$ and $\mathbb{C}^n$ denote a complex number and a complex vector with the length of $n$. $\mathbf{I}$ denotes an identity matrix, while $\mathbf{0}$ denotes a zero matrix. $[\;]^H$, $\Tr(\;)$, $\text{rank}(\;)$ denotes the Hermitian transpose of a matrix, the trace of a matrix and the rank of a matrix. $|\;|$ represents the absolute value. $\succeq$ stands for semi-definite. $\mathcal{CN}(0,\sigma^2)$ denotes standard complex Gaussian distribution with zero mean and variance $\sigma^2$.
\end{comment}

\section{System Model}

%\subsection{Signal Model}
\begin{figure}[!t]
    \centering        
    \includegraphics[width=70mm]{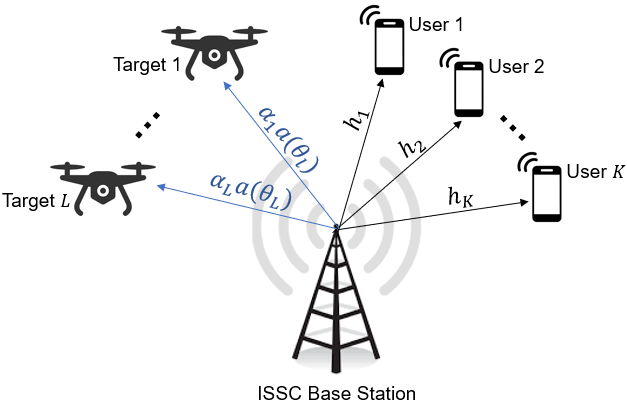}
    \caption{An ISSC system model with $L$ passive targets and $K$   communication users.}\vspace{-0.5em}
    \label{System model}
\end{figure}

As illustrated in Fig. \ref{System model}, we consider an ISSC system with a BS equipped with a uniform linear array (ULA) of $N$ antennas. The antennas are co-located for target detection and downlink semantic communication. The BS communicates with $K$ cellular users (CUs), and each CU $k \in \mathcal{K}$ is equipped with a single antenna. Meanwhile, the BS aims to identify $L$ point-like targets, and each target $l \in \mathcal{L}$ is potentially a passive eavesdropper. 

The received signal at the $k$-th CU can be characterized as follows:
\begin{equation}\label{eq1}
y_k = \mathbf{h}_k^H \mathbf{x} + n_k,
\end{equation}
where $\mathbf{h}_k\in\mathbb{C}^N$ is the channel vector for the $k$-th CU, $\mathbf{x}\in\mathbb{C}^N$ is the transmitted signal, and $n_k \sim \mathcal{CN}(0,\sigma^2_c)$ is the communication noise for the $k$-th CU. On the target side, the received signal can be formulated by
\begin{equation}\label{eq2}
y_l = \alpha_l \mathbf{a}^H(\theta_l) \mathbf{x} + n_l = \mathbf{h}_l^H \mathbf{x} + n_l,
\end{equation}
where $\alpha_l$ and $n_l \sim \mathcal{CN}(0,\sigma^2_r)$ are the path-loss coefficient and the sensing noise for the $l$-th target, respectively. The steering vector is denoted by $\mathbf{a}(\theta_l) \in \mathbb{C}^N$ with $\theta_l$ being the direction of the $l$-th target, $|\theta_l| \leq\frac{\pi}{2}, \forall l$, which can be formulated by 
\begin{equation}\label{eq3}
    \mathbf{a}^H(\theta_l) =  
    \begin{bmatrix}
    1 & e^{j2\pi \frac{d}{\lambda} \sin(\theta_l)} & \cdots & e^{j2\pi (N-1) \frac{d}{\lambda} \sin(\theta_l)}
    \end{bmatrix}.
\end{equation}
In \eqref{eq3}, $\lambda$ is the wavelength and $d$ is the distance between any two adjacent antennas. 
The echo signal received at the BS can be formulated by
\begin{equation}\label{eq4}
    \mathbf{\hat{y}}_l = \beta_l \mathbf{a}(\theta_l) \mathbf{a}^H(\theta_l) \mathbf{x} + \mathbf{n}_l,
\end{equation}
where $\beta_l$ and $\mathbf{n}_l \in \mathbb{C}^{N}$ are the path loss coefficient and noise vector, respectively.

In the considered ISSC system, both semantic communication signals and sensing signals are simultaneously transmitted. Thus, the transmitted signal $\mathbf{x}$ at the BS can be formulated as
\begin{equation}\label{eq5}
    \mathbf{x} = \mathbf{W} \mathbf{c} + \mathbf{R} \mathbf{z},
\end{equation}
where $\mathbf{W} \in \mathbb{C}^{N \times K}$ denotes the precoding matrix and $\mathbf{c} = [c_1, c_2, c_k, \dotsc, c_K] \in \mathbb{C}^{K}$ denotes the semantic message. The semantic message $c_k$ is generated from the conventional communication message $m_k$, i.e., $c_k = f(m_k)$. Here, $f(\cdot)$ is a function that maps a conventional communication message to a semantic message. For instance, $f(\cdot)$ means a neural network with embedded large language model\cite{xie2021deep}, where the semantic message is the output of the neural network. After that, the semantic message is encoded via channel coding before transmission. Another example is to consider $f(\cdot)$ as an autoencoder that achieves joint source channel coding\cite{xie2021deep}. In this paper, we assume that the semantic message is extracted via an autoencoder.

Moreover, in \eqref{eq5}, $\mathbf{R} \in \mathbb{C}^{N \times L}$ represents the radar beamforming matrix and $\mathbf{z} = [z_1, z_2, z_l, \dotsc, z_L] \in \mathbb{C}^{L}$ is the sensing signal. Without the loss of generality, we make the following assumptions:
\begin{enumerate}[i).]
    \item There is no correlation between the confidential message and the radar signal \cite{liuxiang2020joint}. As a result, we have 
    \begin{equation}\label{eq6}
        \mathbb{E}(\mathbf{c} (\mathbf{z})^H) = \mathbf{0}_{K \times L}.
    \end{equation}
    \item There is no correlation between the confidential messages for different CUs. Specifically, this assumption can be stated as follows:
    \begin{equation}\label{eq7}
        \mathbb{E}(\mathbf{c}(\mathbf{c})^H) = \mathbf{I}_{K}.
    \end{equation}
    \item There is no correlation between the sensing signals for different targets. In this case, the assumption can be expressed as follows:
    \begin{equation}\label{eq8}
        \mathbb{E}(\mathbf{z}(\mathbf{z})^H) = \mathbf{I}_{L}.
    \end{equation}
\end{enumerate}

As a result, the covariance matrix of the transmit waveform can be derived as follows:
\begin{equation}\label{eq9}
    \mathbf{R}_x = \mathbb{E}[\mathbf{x}\mathbf{x}^H] = \sum_{k=1}^K \mathbf{W}_k + \sum_{l=1}^L \mathbf{R}_l,
\end{equation}
where $\mathbf{W}_k \equiv \mathbf{w}_k \mathbf{w}_k^H$ and $\mathbf{R}_l \equiv \mathbf{r}_l \mathbf{r}_l^H$. Hence, as given in \cite{fuhrmann2008transmit}, the beampattern can be described by :
\begin{equation}\label{eq10}
    p(\phi) = \mathbf{a}^H(\phi) \mathbf{R}_x \mathbf{a}(\phi),
\end{equation}
where $|\phi| \leq \frac{\pi}{2}$ represents a generic angle covering a fine grid of angles. In this regard, optimizing the transmit beampattern is equivalent to optimizing the covariance matrix.

%\subsection{Performance Indicators}
\subsection{Semantic Communication}
We define the semantic transmission rate as the number of bits received by the users after de-extracting the semantic information, hence the formulation is given by
\begin{equation}\label{eq11}
    S_k = \frac{1}{\rho_k} \log_2(1+\gamma_k),
\end{equation}
where $\rho_k=\frac{\text{len}(c_k)}{\text{len}(m_k)}, 0\leq \rho_k \leq 1$ is the semantic extraction ratio and $\text{len}(\cdot)$ means the number of bits of a message. $\gamma_k$ is SINR for the $k$-th CU which is given by
\begin{equation}\label{eq12}
\footnotesize
\gamma_k= \frac{\Tr(\mathbf{h}_k \mathbf{h}_k^H \mathbf{W}_k)}{\Tr(\mathbf{h}_k \mathbf{h}_k^H \sum^K_{k'=1, k' \neq k}\mathbf{W}_{k'}) +  \Tr(\mathbf{h}_k \mathbf{h}_k^H \sum_{l=1}^L\mathbf{R}_l) + \sigma^2_c}.
\end{equation}

Furthermore, in semantic communication, it is important to consider the similarity between the original message $m_k$ and the extracted message $c_k$. The BLEU score, as defined in \cite{papineni2002bleu}, measures the text message similarity, whose function is given as follows:
\begin{equation*}
\text{BP} = 
    \begin{cases}
    1,& \text{if } \text{len}(c) > \text{len}(m),\\
    e^{1-\frac{\text{len}(m)}{\text{len}(c)}},              & \text{if } \text{len}(c) \leq \text{len}(m).
\end{cases}
\end{equation*}
\begin{equation*}
    \text{BLEU} = \text{BP} \exp(\sum_{g=1}^G w_g \log p_g),
\end{equation*}
where $w_g$ defines the weight of the g-grams and $p_g$ is the g-grams precision score. In our model, the length of $c$ is always less than or equal to the length of $m$. Together with the definition of $\rho_k=\frac{\text{len}(c_k)}{\text{len}(m_k)}$, we can rewrite the BLEU score for the $k$-th CU as:
\begin{equation}\label{eq13}
    \text{BLEU}_k = e^{1-\frac{1}{\rho_k}} \exp(\sum_{g=1}^G w_{g,k} \log p_{g,k}).
\end{equation}

When designing the system, we need to fix the value of $p_{g,k}$ and the lower bound of $\text{BLEU}_k$. The reasons are
\begin{enumerate}[i.]
    \item When optimizing the extraction ratio $\rho_k$, we need to make sure that each user can accurately decode the message.
    \item When $\text{BLEU}_k$ is extremely low, it means the extraction workload for the BS is huge, hence leading to a very high computation cost. 
\end{enumerate}

\begin{lemma}
By choosing an appropriate lower bound value for all $\text{BLEU}_k$ and appropriate values of $p_{g,k}$ for each CU, we can obtain the lower bound of $\rho_k$ as: \begin{equation}\label{eq14}
    \rho_k \geq \frac{1}{1 - \ln Q + \sum_{g=1}^G w_{g,k} \log p_{g,k}}.
\end{equation}
\end{lemma}
In \eqref{eq14}, $Q$ is the lower bound of the all $\text{BLEU}_k$ and $\ln Q \leq \sum_{g=1}^G w_{g,k} \log p_{g,k}$. Additionally, the values of $w_{g,k}$ and $p_{g,k}$ depend on each CU.

\begin{proof}
    See Appendix \ref{appendix1}
\end{proof}

\subsection{Semantic Security}
To measure the level of security, we define the semantic secrecy rate (SSR) of user $k$ as the difference between the semantic transmission rate of user $k$ and the most robust semantic transmission rate of the eavesdropper. To give the formulation, we first need to define the SNR between the BS and $l$-th eavesdropper about the confidential message for the $k$-th CU:
\begin{equation}\label{eq15}
    \Gamma_{l | k}  = \frac{\Tr(\mathbf{h}_l \mathbf{h}_l^H \mathbf{W}_k) }{\Tr(\mathbf{h}_l \mathbf{h}_l^H \sum_{l=1}^L \mathbf{R}_l)  +  \sigma^2_r}.
\end{equation}

Hence, the semantic transmission rate of the $l$-th eavesdropper related to the $k$-th CU is given by 
\begin{equation}\label{eq16}
    S_{l | k} = \frac{1}{\rho_k} \log_2 (1+\Gamma_{l | k}).
\end{equation}

In this way, the worst-case SSR of the $k$-th CU among all possible eavesdroppers can be formulated by
\begin{equation}\label{eq17}
    SSR_{k} = \min_{l \in L} [S_k - S_{l | k}]^+,
\end{equation}
where $[v]^+$ means $\max(0,v)$.

\subsection{Power Budget}

Extracting semantic information from a conventional message is heavily based on machine learning methods. Therefore, we must consider the computation power in the overall power budget. Following the results from \cite{yang2023energy, zhao2023semantic}, the segmented semantic extraction ratio has an exponential relationship with the computation power: when the semantic extraction ratio is closer to zero, the gradient of the function is steeper. However, the function proposed in \cite{zhao2023semantic} depends on many hyper-parameters, such as the interval and the gradient of each segment. To avoid manually defining all the hyper-parameters, we instead use a natural logarithm function to calculate the computation power:
\begin{equation}\label{eq18}
    P_{\text{comp}} =  \sum_{k=1}^K F\ln(1/\rho_k),
\end{equation}
where F is a coefficient that converts a magnitude to its power. On the other hand, the communication and sensing energy consumption is given by
\begin{equation}\label{eq19}
    \Tr(\sum_{k=1}^K \mathbf{W}_k + \sum_{l=1}^L \mathbf{R}_l) = P_{\text{c\&s}}.
\end{equation}

Additionally, the overall power consumption is limited to the power budget:
\begin{equation}\label{eq20}
    P_{\text{comp}} + P_{\text{c\&s}} \leq P_t,
\end{equation}
with $P_t$ being the power budget. 

\subsection{Sensing Performance}
During the design of a sensing-only system, we obtain the radar beamforming matrix $\mathbf{R}_d$, which represents the optimal beam pattern for sensing only. This design can be formulated by maximizing the beams directed toward the targets. As part of this optimization problem, it is necessary to make sure there is no cross-correlation between different targets while also guaranteeing the transmit power budget. Therefore, the maximization problem is given by
\begin{subequations}\label{eq188}
    \begin{align}
        \max_{t,\mathbf{R}_d}\quad & t \label{eq188a}\\
        \text{s.t.} \quad &\mathbf{a}^H(\theta_l) \mathbf{R}_d \mathbf{a}(\theta_l)- \mathbf{a}^H(\theta_m) \mathbf{R}_d \mathbf{a}(\theta_m) \geq t, \forall \theta_l, \forall \theta_m, \label{eq188b}\\
        &\mathbf{a}^H(\theta_l) \mathbf{R}_d \mathbf{a}(\theta_{l'}) = 0, \quad l' \neq l, \quad \forall l,  \label{eq188c}\\
        & \Tr(\mathbf{R}_d) \leq P_t, \label{eq188d}\\
        & \mathbf{R}_d \succeq 0, \mathbf{R}_d = \mathbf{R}_d^H, \label{eq188e}
    \end{align}
\end{subequations}
where $t$ is a threshold value that we wish to maximize, $\theta_l$ represents the angle of the $l^{th}$ target and $ \theta_m \in \Omega$ represents an angle in the sidelobe region. Optimization problem \eqref{eq188} in a pure sensing system is a standard convex optimization problem and thus can be efficiently solved by numerical tools. 

In order to guarantee the sensing performance of the ISSC system, we must guarantee that the mismatch between $\sum_{k=1}^K\mathbf{W}_k+\sum_{l=1}^L\mathbf{R}_l$ and $\mathbf{R}_d$ should be less than a pre-defined threshold $\xi$, i.e., 
\begin{equation}\label{sensing_perf}
    |\mathbf{R}_d - (\sum_{k=1}^K \mathbf{W}_k + \sum_{l=1}^L \mathbf{R}_l)|^2 \leq \xi.
\end{equation}

\section{Beamforming design for Integrated sensing and semantic communication}
In this section, we assume the channel of all CUs and the locations of all targets are available at the BS. 
\subsection{Problem Formulation}
To design secure ISSC beamforming and to optimize the semantic extraction ratio, our goal is to maximize the sum semantic secrecy rate. The constraints of this optimization problem include the boundary of the extraction ratio $\rho_k$ as given in \eqref{eq14}. Next, to guarantee the minimum quality of service for each CU, we require the semantic transmission rate for each CU to be above a pre-defined value $\varsigma$. Additionally, the transmit power budget as defined in \eqref{eq20}, the beampattern mismatch as given in \eqref{eq188}, and the matrix positive semi-definite property need to be satisfied. Finally, to ensure the achievable semantic transmission rate can be attained by using single stream transmit beamforming for each CU, a rank-one constraint is added. Note that if the rank-one constraint is not in place, more sophisticated transceiver schemes are needed \cite{wang2014outage}. 

As a result, the optimization problem can be formulated as follows:
\begin{subequations}\label{eq26}
\begin{align}
    \max_{\mathbf{W}_k, \mathbf{R}_l, \rho_k}\; &  \sum_{k=1}^K SSR_k \label{eq26a}\\
    \text{s.t.} \quad & \frac{1}{1 - \ln Q + \sum_{g=1}^G w_{g,k} \log p_{g,k}} \leq \rho_k \leq 1, \forall k,\label{eq26b}\\
    & S_k \geq \varsigma, \forall k, \label{eq26c}\\
    &  P_{\text{comp}} + P_{\text{c\&s}} \leq P_t,\label{eq26d}\\
    & |\mathbf{R}_d - (\sum_{k=1}^K \mathbf{W}_k + \sum_{l=1}^L \mathbf{R}_l)|^2 \leq \xi,\label{eq26new}\\
    &\mathbf{W}_k \succeq 0, \mathbf{W}_k = \mathbf{W}_k^H, \quad \forall k\label{eq26e},\\
    & \mathbf{R}_l \succeq 0, \mathbf{R}_l = \mathbf{R}_l^H, \forall l\label{eq26f},\\
    & \text{rank}(\mathbf{W}_k) = 1, \quad \forall k.\label{eq26g}
\end{align}
\end{subequations}

The worst-case SSR in \eqref{eq26a} can be modified to:
\begin{equation}\label{eq27}
    SSR_{k} =  [S_k - \max_{l\in L} S_{l | k}]^+.
\end{equation}

\subsection{Algorithm Design}
Through introducing auxiliary variables $\lambda_k, \forall k$, together with \eqref{eq27}, we can transform the first part of \eqref{eq26a} into the following form:
\begin{equation}\label{eq28}
    \max_{\mathbf{W}_k, \mathbf{R}_l, \rho_k, \lambda_{k}} \sum_{k=1}^K  (S_k - \frac{1}{\rho_k} \log_2(1+\lambda_{k})),
\end{equation}
with an additional constraint of
\begin{equation}\label{eq29}
    \Gamma_{l | k} \leq \lambda_{k}, \forall k, \forall l.
\end{equation}

However, \eqref{eq28} is a non-concave function due to the logarithm terms in the equation. We can apply first-order Taylor expansion on $S_k$, which leads to:
\begin{equation}\label{eq30}
\begin{aligned}
&S_k \\
& = \frac{1}{\rho_k} \log_2(\frac{\mathbf{h}_k^H \sum_{k=1}^K \mathbf{W}_k \mathbf{h}_k + \mathbf{h}_k^H \sum_{l=1}^L \mathbf{R}_l \mathbf{h}_k + \sigma_c^2 }{\mathbf{h}_k^H \sum_{k'=1, k'\neq k}^K \mathbf{W}_{k'} \mathbf{h}_k + \mathbf{h}_k^H \sum_{l=1}^L \mathbf{R}_l \mathbf{h}_k + \sigma_c^2 }) \\
& = \frac{1}{\rho_k} \Big( \log_2(A_k) - \log_2(B_k) \Big)\\
& = \frac{1}{\rho_k} \Big( \log_2(A_k) - \log_2(B_k^i) - \frac{1}{B_k^i \ln(2)} (B_k - B_k^i) \Big).
\end{aligned}
\end{equation}
As a result, the maximization of the concave expression $S_k$ in \eqref{eq30} falls in the convex optimization.

Similarly, by applying first-order Taylor expansion on $-\frac{1}{\rho_k}\log_2(1+\lambda_{k})$, we can further have:
\begin{equation}\label{eq31}
    \begin{aligned}
        &-\frac{1}{\rho_k}\Big( \log_2(1 + \lambda_{k}) \Big) = -\frac{1}{\rho_k} \Big( \log_2(C_{k}) \Big)\\
        & = -\frac{1}{\rho_k} \Big( \log_2(C_{k}^i) + \frac{1}{C_{k,l}^i \ln(2)} (C_{k} - C_{k}^i) \Big).\\
    \end{aligned}
\end{equation}

In \eqref{eq30} and \eqref{eq31}, $A_k$, $B_k$, $C_{k}$ are variables and $B_k^i$, $C_{k}^i$ stands for the value of $B_k$, $C_{k}$ in the $i$-th iteration, which can be calculated by
\begin{equation}
    \begin{aligned}
        B_k^i &= \mathbf{h}_k^H \sum_{k'=1, k'\neq k}^K \mathbf{W}^i_{k'} \mathbf{h}_k + \mathbf{h}_k^H \sum_{l=1}^L \mathbf{R}^i_l \mathbf{h}_k + \sigma_c^2 \\
        C_k^i & = 1 + \lambda^i_k,  
    \end{aligned}
\end{equation}
where $W^i_{k'}$, $R_l^i$ and $\lambda^i_k$ are the values of $W_{k'}$, $R_l$ and $\lambda_k$ in the $i$-th iteration.

Furthermore, by dropping the rank-one constraint using SDR, problem \eqref{eq26} can be transformed into the following form
\begin{subequations}\label{eq36} 
\begin{align}
    \max_{\psi} &  \sum_{k=1}^K  \frac{1}{\rho_k} \Big( \log_2(A_k) - \log_2(B_k^i) - \frac{1}{B_k^i \ln(2)} (B_k - B_k^i) \nonumber\\
    & - \log_2(C_{k}^i) - \frac{1}{C_{k}^i \ln(2)} (C_{k} - C_{k}^i)\Big) ,\label{eq36a}\\
    \text{s.t.} \quad & \frac{1}{1 - \ln Q + \sum_{g=1}^G w_{g,k} \log p_{g,k}} \leq \rho_k \leq 1, \forall k,\label{eq36b}\\
    & \frac{1}{\rho_k} \Big( \log_2(A_k) - \log_2(B_k^i) - \frac{1}{B_k^i \ln(2)} \nonumber \\  &(B_k - B_k^i) \Big)\geq \varsigma, \forall k, \label{eq36c}\\
    & \Gamma_{l | k} \leq \lambda_{k}, \forall k, \forall l, \\
    &  P_{\text{comp}} + P_{\text{c\&s}} \leq P_t, \label{eq36f}\\
    & |\mathbf{R}_d - (\sum_{k=1}^K \mathbf{W}_k + \sum_{l=1}^L \mathbf{R}_l)|^2 \leq \xi,\label{eq36new}\\
    &\mathbf{W}_k \succeq 0, \mathbf{W}_k = \mathbf{W}_k^H, \forall k,\label{eq36g}\\
    & \mathbf{R}_l \succeq 0, \mathbf{R}_l = \mathbf{R}_l^H, \forall l,\label{eq36h}
\end{align}
\end{subequations}
where $\psi = [\mathbf{W}_k, \mathbf{R}_l, \rho_k, \lambda_k]$. To solve \eqref{eq36}, we consider using alternating optimization:
\subsubsection{Step 1} 
%Initialize the values of $\rho_k$, $\mathbf{W}_k^i$, $\mathbf{R}_l^i$, $\lambda_{k}$, then solve for $\mathbf{W}_k$, $\mathbf{R}_k$:
With given $\rho_k$ and $ \lambda_k$ in problem \eqref{eq36}, the joint sensing and communication beamforming optimization problem can be given by
\begin{subequations}\label{eq37}
\begin{align}
    \max_{\mathbf{W}_k, \mathbf{R}_l}\; &  \sum_{k=1}^K \frac{1}{\rho_k} \Big( \log_2(A_k) - \log_2(B_k^i) - \frac{1}{B_k^i \ln(2)} \nonumber\\ &(B_k - B_k^i) - \log_2(1+\lambda_{k}) \Big) \label{eq37a}\\
    \text{s.t.} \quad & \frac{1}{\rho_k} \Big( \log_2(A_k) - \log_2(B_k^i) - \frac{1}{B_k^i \ln(2)} \nonumber \\  &(B_k - B_k^i) \Big)\geq \varsigma, \forall k, \label{eq37b}\\
    & \Tr(\mathbf{h}_l \mathbf{h}_l^H \mathbf{W}_k) - \nonumber \\ &\lambda_k (\Tr(\mathbf{h}_l \mathbf{h}_l^H \sum_{l=1}^L \mathbf{R}_l)  +  \sigma^2_r) \leq 0, \forall k, \forall l, \\
    &  P_{\text{comp}} + P_{\text{c\&s}} \leq P_t\label{eq37e}\\
    & |\mathbf{R}_d - (\sum_{k=1}^K \mathbf{W}_k + \sum_{l=1}^L \mathbf{R}_l)|^2 \leq \xi, \\
    &\mathbf{W}_k \succeq 0, \mathbf{W}_k = \mathbf{W}_k^H, \quad \forall k\label{eq37f}\\
    & \mathbf{R}_l \succeq 0, \mathbf{R}_l = \mathbf{R}_l^H, \forall l,\label{eq37g}
\end{align}
\end{subequations}
which is convex and can be effectively solved via the standard convex optimization tool.

\subsubsection{Step 2} With fixed $\rho_k$ as well as the obtained values of $\mathbf{W}_k$ and $\mathbf{R}_k$ from Step 1, the optimization of  $\lambda_{k}$ is 
\begin{subequations}\label{eq38} 
\begin{align}
    \max_{\lambda_k}\; &  \sum_{k=1}^K \frac{1}{\rho_k} \Big( \log_2(A_k) - \log_2(B_k) \nonumber\\ 
    &-\log_2(C_{k}^i) 
    -\frac{1}{C_{k}^i \ln(2)} (C_{k} - C_{k}^i) \Big) \label{eq38a}\\
    \text{s.t.} \quad & \lambda_k \geq \Gamma_{l | k}, \forall k, \forall l,
\end{align}
\end{subequations}
where $A_k$, $B_k$ and $\Gamma_{l | k}$ are known values. $C_k^i$ is updated iteratively and $C_k = 1 + \lambda_k$ is the variable. Problem \eqref{eq38} is convex and can be effectively solved.

\subsubsection{Step 3} With given $\mathbf{W}_k$, $\mathbf{R}_k$, $\lambda_k$, the optimization of semantic compression ratio $\rho_k$ is
\begin{subequations}\label{eq39} 
\begin{align}
    \max_{\rho_k} &  \sum_{k=1}^K  \frac{1}{\rho_k} \Big( \log_2(A_k) - \log_2(B_k) - \log_2(C_{k}) \Big) \label{eq39a}\\
    \text{s.t.} \quad & \frac{1}{1 - \ln Q + \sum_{g=1}^G w_{g,k} \log p_{g,k}} \leq \rho_k \leq 1, \forall k \label{eq39b}\\
    & \frac{1}{\rho_k} \Big( \log_2(A_k) - \log_2(B_k)\Big)\geq \varsigma, \forall k \label{eq39v}\\
    &  P_{\text{comp}} \leq P_t - P_{\text{c\&s}} \label{eq39d}
\end{align}
\end{subequations}
which is also convex and can be solved through the dual method.

\subsubsection{Step 4}
As a last step, we use Gaussian randomisation to recover rank-one solutions. 

The detailed procedures for solving problem \eqref{eq36} are shown in Algorithm 1.

\begin{algorithm}
\caption{Iterative Sensing, Communication, and Semantic Optimization Algorithm}\label{alg:1}
\begin{algorithmic}[1]
\STATE Initialize $\rho_k$, $\mathbf{W}_k^i$, $\mathbf{R}_k^i$, $\lambda_k$, $B_k^i$ and $C_k^i$.
\REPEAT
\REPEAT
    \STATE Solve \eqref{eq37}.
    \STATE Update $\mathbf{W}_k^{i+1}$ and $\mathbf{R}_k^{i+1}$.
\UNTIL{$|\mathbf{W}^{i+1} - \mathbf{W}^i| \leq \varrho_1$, $|\mathbf{R}^{i+1} - \mathbf{R}^i| \leq \varrho_2$.}
\REPEAT
    \STATE Solve \eqref{eq38}.
    \STATE Update $C_k^{i+1}$.
\UNTIL{ $|C_k^{i+1} - C_k^i| \leq \varrho_3$.}
    \STATE Solve \eqref{eq39} and update $\rho_k$.
\UNTIL{converge}
\STATE Apply Gaussian randomisation.
\end{algorithmic}
\end{algorithm}

\section{Numerical results}

During our simulation, we assume the BS is equipped with ULA that has half-wavelength spacing. The number of antennas is 18. The locations of targets are at $[-35^\circ, 5^\circ, 45^\circ]$ and the locations of CUs are at $[-30^\circ, 20^\circ]$. Path loss coefficients are generated randomly between $0.001$ and $0.01$. The value of $\varsigma$ is equal to 1 and $\xi$ is equal to 5. The average noise power is -60 dBm. The value of $F$ in \eqref{eq18} is 10. The power budget varies from 5 dBm to 25 dBm. For comparison, the first benchmark we use is by setting $\rho_k = 1, \forall k$ and never updating it. This means there is no semantic extraction. 
The second benchmark is \cite{9199556}, where the SINR threshold upper bound is 25.

\begin{figure}
    \centering
    \includegraphics[width=70mm]{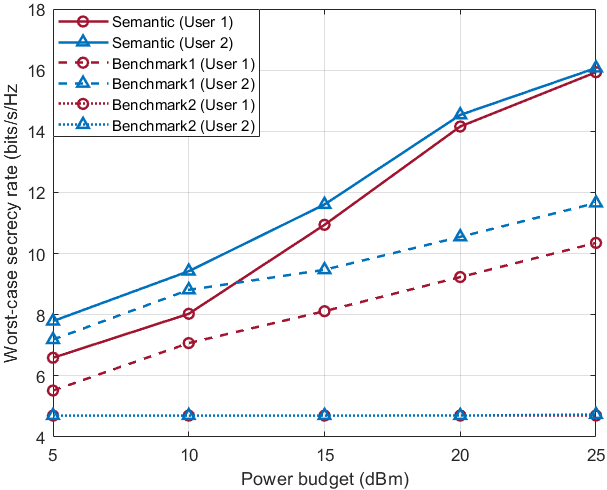}
    \caption{Worst-case semantic secrecy rate for different users.}
    \label{fig1}
\end{figure}

\begin{figure}
    \centering
    \includegraphics[width=70mm]{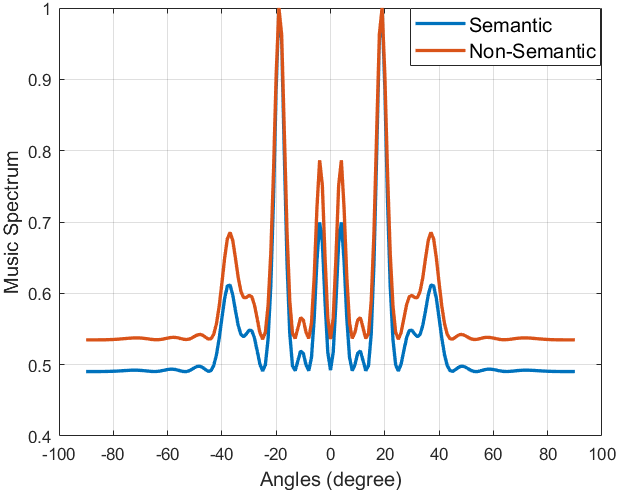}
    \caption{MUSIC Spectrum}
    \label{fig2}
\end{figure}

%By analyzing the worst-case SSR achieved by each user, we can gain a deeper insight into the algorithm's performance. 
Fig. \ref{fig1} shows that when semantic extraction is involved in the system, the worst-case SSR improves significantly compared to benchmark 1 and benchmark 2. Furthermore, user 1 achieves a slightly lower worst-case SSR than user 2. This is because we assume user 1 is less capable of recovering the semantic message. Therefore, the lower bound of $\rho_{k=1}$ is higher than that of $\rho_{k=2}$, which leads to a lower worst-case SSR for user 1.
In Fig. \ref{fig2}, we evaluate the sensing performance using the Multiple SIgnal Classification (MUSIC) spectrum. In both our algorithm and the benchmarks, we are able to detect peaks at the locations of targets, which indicates that the radar is capable of detecting all targets accurately. Thus, our algorithm guarantees the same sensing performance as the benchmarks while providing higher security.

\section{Conclusion}
This paper provided a unified framework of  integrated sensing and semantic communication. A joint design of transmit beamforming and semantic extraction ratio was presented. By using alternating optimization, we proposed an iterative algorithm to optimize the sum of the semantic secrecy rate for all users. Based on our simulation results, we found that our design provides higher security than the existing ISAC frameworks while guaranteeing the same sensing performance.

\appendices

\section{Proof of Lemma 1} \label{appendix1}
\setcounter{equation}{0}
\numberwithin{equation}{section}
By taking the natural logarithm on both sides of \eqref{eq13} and setting the lower bound of $\text{BLEU}_k$, we have:
\begin{equation}
    \ln(\text{BLEU}_k) = 1 - \frac{1}{\rho_k} + \sum_{g=1}^G w_{g,k} \log p_{g,k} \geq \ln Q,
\end{equation}
with some simple algebra operations, we have:
\begin{equation}
    \rho_k \geq \frac{1}{1 - \ln Q + \sum_{g=1}^G w_{g,k} \log p_{g,k}}.
\end{equation}

\bibliographystyle{ieeetr}
\bibliography{bib}

\end{document}